# Assessment of Natural Radioactivity Concentration and Radiological Risk in Tanjung Enim's Coal Mine, South Sumatra Indonesia

Muhammad Farhan Ramadhany ( ✉ farhanramadhany@mail.ugm.ac.id )
  Department of Nuclear Engineering and Engineering Physics, Faculty of Engineering, Gadjah Mada University    https://orcid.org/0000-0003-3183-9063

Gede Sutresna Wijaya
  National Research and Innovation Agency, Yogyakarta

Anung Muharini
  Department of Nuclear Engineering and Engineering Physics, Faculty of Engineering, Gadjah Mada University

# Assessment of Natural Radioactivity Concentration and Radiological Risk in Tanjung Enim's Coal Mine, South Sumatra Indonesia

Muhammad Farhan Ramadhany[a*], Gede Sutresna Wijaya[b], Anung Muharini[a]

[a] *Department of Nuclear Engineering and Engineering Physics, Faculty of Engineering of Universitas Gadjah Mada, Republic of Indonesia*
[b] *Yogyakarta National Research and Innovation Agency, Republic of Indonesia*

**Abstract** Environmental radioactivity analysis has been carried out to determine the level of environmental radioactivity and the potential radiological hazards at Tanjung Enim's coal mine. Gamma spectroscopy method has been carried out to identify radionuclides and their types of activity. The results of radionuclide concentration are used to determine the radiological hazard index and become input data for the Residual Radioactivity Onsite 7.2 application to determine the dose rate and long-term cancer potential received by workers in coal mines. The results obtained for the average concentration of radionuclide activity in coal samples are $^{226}$Ra, $^{232}$Th, and $^{40}$K are 72.468 Bq/kg, 86.905 Bq/kg, and 1802.049 Bq/kg, respectively. While the soil samples $^{226}$Ra, $^{232}$Th, and $^{40}$K respectively 79.205 Bq/kg, 100.209 Bq/kg, and 1443.275 Bq/kg. The radionuclide concentrations of both samples exceeded the UNSCEAR and worldwide reported averages for coal and soil. The average radiological hazard index for coal samples, namely Ra$_{eq}$, H$_{in}$, and H$_{ex}$, was 335.500 Bq/kg, 1.102, 0.906, respectively. While the soil samples were 333.636 Bq/kg, 1.115, and 0.901, respectively. The index parameter is already lower than the UNSCEAR recommendation except for H$_{in}$, so there is a potential radiological hazard in internal pathways such as respiratory and digestive organs for mining workers. The total annual effective dose rate based on 5 RESRAD-Onsite 7.2 pathways, namely external gamma, inhalation, radon, soil ingestion, and drinking water, is 1.675 mSv/year, exceeding the dose limits determined by ICRP, 1 mSv/year. The ELCR is 6.625×10$^{-3}$ which exceeds the UNSCEAR recommendation, 2.4×10$^{-4}$. Based on the results, it is necessary to intervene in the mining environment of the Tanjung Enim's coal mine.

**Keywords** Coal mine; Environmental radioactivity; Radiological hazard; Gamma spectrometry

## 1. Introduction

In Indonesia, coal is used as the primary material to meet Indonesia's energy needs. Even up to 2016, domestic coal consumption reached 76% of which was used by steam power plants (Haryadi & Suciyanti, 2018). The use of coal manages the high price of fuel oil due to its reduced reserves. The reserves in Indonesia are still relatively abundant and recorded in 2016 was still at 28.46 billion tons with an estimated run-out time of about 68 years (BPPT, 2018). Meanwhile, based on the number of coal reserves in Tanjung Enim, it is recorded that it has mined coal reserves of 3.33 billion tons and resources of 8.17 billion tons (Bukit Asam Press, 2020).

These coal mining and consumption activities produce pollutants that pollute the air and soil, such as carbon monoxide, nitrogen oxides, sulphur gases, and hydrocarbon compounds, followed by the release of radioactive substances into the environment (Güllüdağ et al., 2020; Habib et al., 2019; Makolli et al., 2020; Makudi et al., 2018; Yadav et al., 2020). This is because coal naturally contains several radionuclide elements such as primordial radionuclides, that labelled as naturally occurring radioactive material (NORM), at coal mining sites contains uranium-series radionuclides with uranium parent ($^{238}$U), thorium series ($^{232}$Th), and primordial radionuclides not from series such as potassium ($^{40}$K) (Dumitrescu et al., 2018; Makudi et al., 2018; Monged, 2020). These mining activities can redistribute and enhance the concentration of natural radionuclides to the surrounding environment. Likewise, oil and gas exploration, thermal power generation, and the natural materials processing industry can alter natural radioactivity in different process states. Waste generated by this industry must be handled with care, subject to natural radioactivity levels and under national and international regulations. They are called technologically enhanced natural radioactive materials (TENORM).

In the present work the activity concentrations of $^{226}$Ra, $^{232}$Th, and $^{40}$K radioisotopes in coal and soil samples were determined. All these radionuclides have very long half-lives. The naturally occurring isotope $^{226}$Ra as part of $^{238}$U decay series is the most toxic of radium isotopes (Gössner, 1999; Vukanac et al., 2022). Concentrated radioactive pollutants can be in excavations and landfills, it is feared that they can cause radiological health problems. Thus, they may reach human body through the intake of contaminated water, food, and soil, the inhalation of particulate pollutants, and exposure to external radiation which can cause various diseases, e.g., cell damage, lung and bone cancer (Habib et al., 2019; Munawer, 2018). This is reinforced by the results of epidemiological studies in various countries showing that radon, as a decaying child of the uranium series, and its derivatives cause carcinogenic effects on mining workers. Studies of exposed miners have consistently found an association between radon and lung cancer. According to the World Health Organization (WHO), a statistically significant increase in the risk of lung cancer occurs as a result of prolonged exposure to radon when its concentration is at 100 Bq/m$^3$, and increases by 16% per every 100 Bq/m$^3$ (Wysocka et al., 2019). Workers who continuously work at coal mining sites can have an annual dose exceeding the limit set by ICRP publication 103, with a dose limit value that can be accepted by non-radiation workers who are part of the general public is 1 mSv/year (ICRP, 2007, 2014, 2020).

In this regard, not only from nuclear technology activities but radioactive substances at coal mining sites also contribute to increasing environmental radioactivity and potentially disrupting human health around the mine. Therefore, further research is needed on radionuclide's types and their concentrations in mining minerals. The radioactivity is used to obtain radiological hazard information, such as the hazard index, absorption dose, annual effective dose, and cancer risk from the minerals received by coal-mining workers at the Tanjung Enim's coal mine, South Sumatra, Indonesia. The study's findings will be useful in assessing public radiation doses and monitoring environmental radioactivity. The results of this study are also anticipated to apply to the effective management of radiogenic pollutants.

[*] Corresponding author
e-mail: farhanramadhany@mail.ugm.ac.id (M. F. Ramadhany)

## 2. Materials and Methods

### 2.1. Research Tools and Materials

The tools used in this study consisted of sampling tools, namely shovel, crowbar, hammer, hoe, 1-liter bottle, basket, label, and GPS. Tools for sample preparation, namely, tray, mesh 200, mortar and pastel, spoon, brush, bottle vials, plastic glue, digital scales, oven, and labels. Tools for sample counting gamma are spectrometer systems with an HPGe detector and a vial bottle holder. Software used for analysis are Maestro 7.01 and RESRAD-Onsite 7.2.

The research materials used in this study were three samples of heap cluster soil, three samples of viewpoint cluster soil, three samples of stockpile cluster coal, three samples of mine pit cluster coal, and the IAEA Soil-6 standard source with an activity of $^{226}$Ra 79.90 Bq/kg per January 30, 1983.

### 2.2. Research Procedure

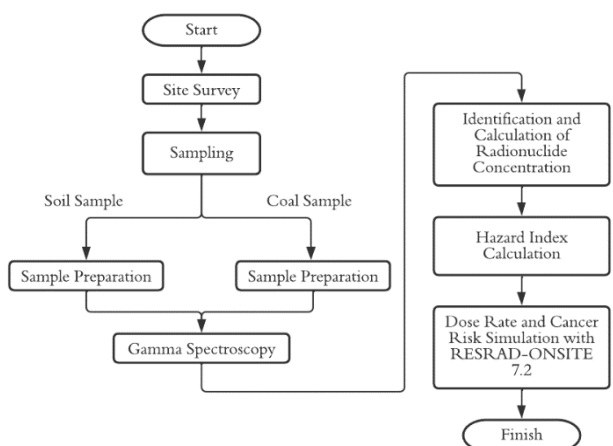

**Figure 1**. Research flow chart consisting of site survey, sampling, sample preparation, gamma spectrometry analysis, and modeling with the RESRAD-ONSITE 7.2 application

#### 2.2.1. Site Survey

The location survey is intended to study and find strategic places and can be used as sites for this research. The survey results will be used to overview the location, area, and activities at the Tanjung Enim's coal mine, South Sumatra, Indonesia, which has operated for more than 70 years. The length of time mining operations leads to a high potential for NORM accumulation so that it has a more dominant influence on the radiological health of coal mining workers.

The results of the site survey at the mine site of concern consist of 4 clusters, namely the TSBC Front active mining pit cluster, the Suban soil stockpile cluster, the coal stockpile cluster, and the TSBC viewpoint cluster and stockpile viewpoint. Pick-up points determine each cluster based on human activities, types of activities, geographical conditions, and permits granted by the company. The details of the survey of sampling locations in each cluster can be seen in Table 1 and Figure 3.

**Table 1.** Coordinates of sampling in various clusters at the Tanjung Enim coal mine, Indonesia

| Cluster Location | Cluster Area (Ha) | Number of Samples | Sample Type | Sample Coordinate, UTM |
|---|---|---|---|---|
| Pit Front TSBC | 20,30 | 3 | Coal | (363100,9583418) (363090,9583440) (363080,9583380) |
| Coal Stockpile | 15,29 | 3 | Coal | (363971,9585857) (363962,9585900) (363967,9585960) |
| Suban Soil Stockpile | 3,10 | 3 | Soil | (364595,9584559) (364574,9584523) (364574,9584529) |
| Viewpoint TSBC | 0,22 | 2 | Soil | (362741,9584894) (362698,9584884) |
| Viewpoint Stockpile | 0,26 | 1 | Soil | (364128,9584298) |

#### 2.2.2. Sample Collection

The sampling method for each cluster is simplified random sampling to make the results more representative of the actual conditions. The samples taken consisted of soil and coal samples. Each cluster has taken three samples. Every sample had a 500 to 1000 ml volume and a depth of 5 to 10 cm. A sampling at these depths because the density of contamination in the first year or two after deposition can usually be determined at ground level. (Barnekow et al., 2019; IAEA, 2004). The location code and coordinates are recorded for each sampling at the location. The sample that has been taken is placed in a plastic container and then given an identity then put into a container for transportation to the next location.

#### 2.2.3. Sample Preparation

Soil and coal sample preparation can follow the flow chart in Figure 2.

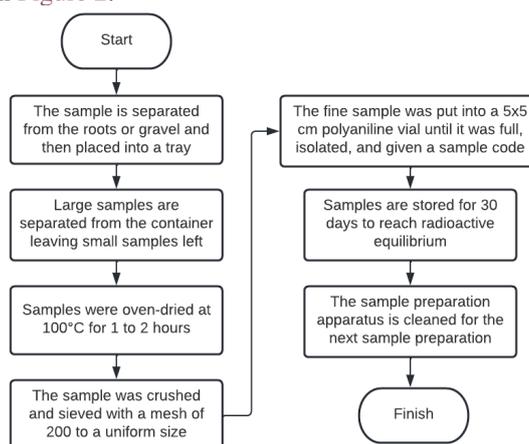

**Figure 2.** Flowchart of sample preparation, consisting of separation of impurities, drying, homogenization and storage

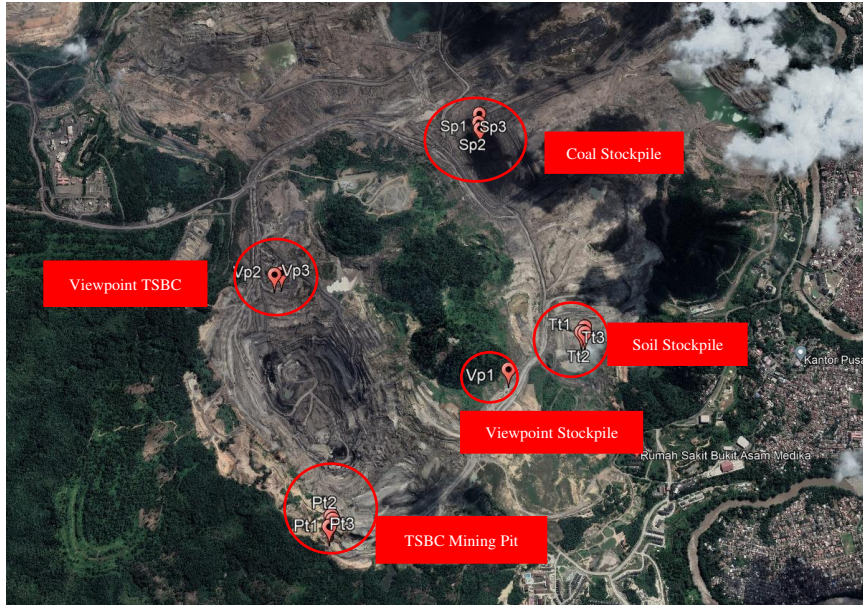

**Figure 3.** Location of sample collection point in the Tanjung Enim's coal mine, Indonesia (Image © 2021 Google)

### 2.2.4. Radiation Counting

Samples that have been stored for 30 days have reached radioactive equilibrium and are ready to be counted. The detector was calibrated by counting standard and background sources for 12 hours. Then the sample was prepared and placed on top of the HPGe detector. Samples counted with 9 hours, 14 hours, and 16 hours. The counting data form of the spectrum of each sample stored in a computer folder.

### 2.2.5. Research Analysis

#### i. Energy Calibration

Calculation of energy calibration can determine the relationship between the channel number and the gamma energy. Each radionuclide has a specific energy, and energy calibration will be used as the basis for qualitative and quantitative research analysis (Knoll, 2010; Tsoulfanidis & Landsberger, 2015). Energy calibration is carried out by counting standard radionuclide sources and standard soil by following the data contained in the certificate.

The energy calibration is carried out by counting the standard source, IAEA Soil 6, for 12 hours and performing energy calibration through a graph of the relationship between channel number and energy. This relationship can be approximated by linear regression.

#### ii. Efficiency Calibration

Efficiency calibration in sample counting is carried out to determine the detector's efficiency at a certain energy level or range. Quantitative radionuclide analysis is carried out based on this efficiency calibration. (Wahyudi et al., 2007). Efficiency calibration includes calculating the efficiency of the semiconductor detector system as a function of energy. It also includes correction factors caused by the intrinsic detector crystal, detector source geometry, the material around the detector and absorption in the source matrix (Dovlete & Povinec, 2004). Efficiency calibration is required for each source-detector combination. After efficiency calibration with a secondary standard, in most cases prepared in the same geometry and matrix as the unknown sample, the sample is counted, usually for 12 hours or more, to satisfy the required statistical uncertainty. The counting period depends on the activity of the sample. The efficiency value for each energy can be calculated using Eq. (1) (BATAN, 2013; Wahyudi & Wilyono, 2006).

$$\varepsilon_\gamma(E_\gamma) = \frac{\frac{C_{st}}{t_{st}} - \frac{C_{bg}}{t_{bg}}}{A_{st} \times m_{st} \times p_\gamma \times f_k} \quad (1)$$

$$f_k = \frac{\mu x}{1 - e^{-\mu x}} \quad (2)$$

$$\mu = \mu_m \times \rho \quad (3)$$

$$\mu_m = 1{,}287 \times E_\gamma^{-0{,}435} \quad (4)$$

with $\varepsilon_\gamma$ is counting efficiency; E is gamma energy (keV); $C_{st}$ is standard source count; $C_{bg}$ is background count; $t_{st}$ is standard source count time (s); $t_{bg}$ is background count time (s), $A_{st}$ is standard source radioactive concentration (Bq/kg); $m_{st}$ is standard source mass (kg); $p_\gamma$ is source yield; fk is absorption factor; $\mu$ is linear attenuation coefficient (cm$^{-1}$); $\mu_m$ is mass attenuation coefficient (g/cm$^2$); $\rho$ is sample density (g/cm$^3$); x is thickness sample (cm).

#### iii. Radioactivity Analysis in Sample

Soil and coal samples were counted at various times, namely 9 hours, 14 hours, and 16 hours. Long counting time so that the radionuclide spectrum of interest in the sample can be distinguished from the background radiation. The resulting spectrum will be analyzed to determine the concentration of radionuclides contained in the sample. Then the calculation of Lower Limit Detection (LLD) is also carried out as a benchmark for the lowest concentration level that can be determined statistically different from the blank at the 99% confidence level. In other words, it is the lowest amount of a substance that can be distinguished from the absence of that substance. To calculate the concentration of each radionuclide and LLD can use Eq. (5) and Eq. (6) (BATAN, 2013; Wahyudi & Wilyono, 2006).

$$A_{spl} = \frac{\frac{C_{spl}}{t_{spl}} - \frac{C_{bg}}{t_{bg}}}{\varepsilon_\gamma \times m_{spl} \times p_\gamma \times f_k} \quad (5)$$

$$LLD\ (Bq) = \frac{4{,}66\sqrt{\frac{C_{bg}}{t_{bg} \times t_{st}}}}{\varepsilon_\gamma \times p_\gamma \times f_k} \quad (6)$$

*iv. Analysis of Radiological Hazard Parameters*

Parameters of radiological hazards posed by mining minerals to workers can be estimated or calculated by determining the hazard index, radiation dose rate, and cancer potential. The hazard index analysis was carried out by calculating the radium equivalent activity ($Ra_{eq}$), the external hazard index ($H_{ex}$), and the internal hazard index ($H_{in}$) using the equations compiled by ICRP in publication 60 and UNSCEAR. (ICRP, 1990; UNSCEAR, 2000).

$$Ra_{eq} = A_{Ra} + \frac{370}{259} \times A_{Th} + \frac{370}{4810} \times \quad (7)$$

$$H_{ex} = \frac{A_{Ra}}{370} + \frac{A_{Th}}{259} + \frac{A_K}{4810} \quad (8)$$

$$H_{in} = \frac{A_{Ra}}{185} + \frac{A_{Th}}{259} + \frac{A_K}{4810} \quad (9)$$

with $A_{Ra}$ is concentration of $^{226}$Ra; $A_{Th}$ is concentration of $^{232}$Th; $A_K$ is concentration of $^{40}$K.

The dose rate and cancer risk were calculated by simulation RESRAD-Onsite 7.2. The parameter is calculated by opening five pathways: direct external radiation, inhalation, radiation exposure, soil ingestion, and drinking water. The data used are based on the type of activity obtained in the enumeration of soil and coal samples and for environmental parameters using standard parameters or by the Tanjung Enim's coal mine conditions.

## 3. Results and Discussion

### 3.1. Gamma Spectrometer Energy Calibration

Energy calibration is carried out by counting the standard source of Soil 6 for 12 hours. Calibration is done by looking at the energy peaks of the radionuclides recorded in the standard source certificate on each spectrum channel. The radionuclides contained in the standard sources are $^{226}$Ra, $^{40}$K, $^{137}$Cs, $^{90}$Sr, $^{239}$Pu, and $^{240}$Pu. The calibration results are presented in Table 2 and Figure 4.

**Table 2.** Position of radiant energy with respect to channel number based on IAEA-soil standard 6 . source spectrum

| Energy (keV) | Channel |
|---|---|
| 51.6 | 296 |
| 295.2 | 1627 |
| 351.9 | 1939 |
| 609.3 | 3345 |
| 661.7 | 3631 |
| 1460.8 | 8000 |
| 1761 | 9659 |

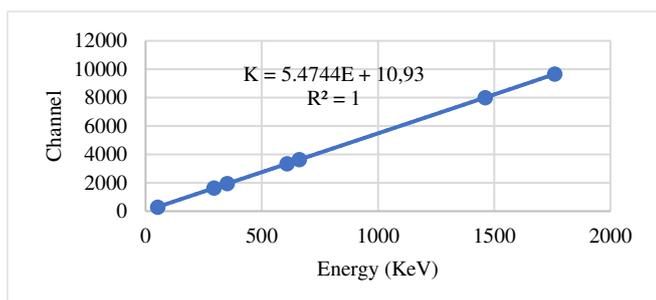

**Figure 4.** Energy calibration results showing the correlation between energy and spectrometric channel numbers

### 3.2. Counting Efficiency Calibration

Efficiency calibration using standard IAEA soil 6 sources containing radionuclides such as $^{226}$Ra, which has been in secular equilibrium, i.e. when the concentration or activity of parent and offspring is the same because the half-life of the parent is much longer than the half-life of the offspring. The activity of the type $^{226}$Ra in soil 6 was recorded at 79.90 Bq/kg as of January 30, 1983. This made the measurement of the activity of $^{226}$Ra possible by measuring the activity of its entire daughter nuclides, such as $^{214}$Pb (295.2 keV and 351.9 keV) and $^{214}$Bi (609.3 keV energy). Determination of $^{226}$Ra activity cannot by looking at the spectrum at the energy of $^{226}$Ra itself, which is 186.211 keV, due to its low energy yield (3.64%) and the intervention of decay energy of $^{235}$U, 185.75 keV, making the activity measurement inaccurate. The following is the result of calculating the efficiency of $^{226}$Ra counting on soil 6 present in Table 3 and Figure 5.

**Table 3.** Counting efficiency on element $^{226}$Ra based on IAEA-soil 6 standard source counting using gamma spectroscopy system

| Parent nuclide | Daughter nuclide | Energy (keV) | Yield (%) | Efficiency |
|---|---|---|---|---|
| $^{226}$Ra | $^{214}$Pb | 295.2 | 18.42 | 0,0282 |
| | $^{214}$Pb | 351.9 | 35.60 | 0,0248 |
| | $^{214}$Bi | 609.3 | 45.49 | 0,0155 |

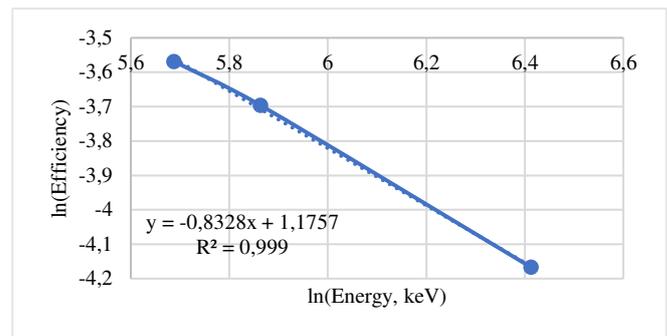

**Figure 5.** Counting efficiency calibration results for each energy range

Since efficiency is a function of energy, the counting efficiency of other radionuclides such as $^{232}$Th and $^{40}$K can be calculated. The efficiency of radionuclide $^{232}$Th can be determined from the decay of $^{212}$Pb at 238.6 keV, $^{228}$Ac at 338.3 and 911.1 keV, and $^{208}$Tl at 583.2 keV. The selection of decayed daughter nuclides to determine the efficiency of $^{232}$Th was based on the relatively shorter half-life of the daughter radionuclides so that the offspring's activity would increase more quickly to reach equilibrium. In addition, it is also based on the yield of the daughter radionuclides, which is large enough so that the results of the counting will be significantly different from the background count. Meanwhile, the efficiency of the $^{40}$K radionuclide is determined directly by its decay energy, which is 1460,8 keV. The results of the calculation of the counting efficiency can be seen in Table 4.

**Table 4.** The results of the counting efficiency of each energy emitted by the radionuclides analyzed in this study

| Parent nuclide | Daughter nuclide | Energy (keV) | Yield (%) | Efficiency |
|---|---|---|---|---|
| $^{226}$Ra | $^{214}$Pb | 295.2 | 18.42 | 0.0282 |
| | $^{214}$Pb | 351.9 | 35.60 | 0.0248 |
| | $^{214}$Bi | 609.3 | 45.49 | 0.0155 |
| $^{232}$Th | $^{212}$Pb | 238.6 | 43.60 | 0.0339 |
| | $^{228}$Ac | 338.3 | 11.27 | 0.0254 |
| | $^{228}$Ac | 911.1 | 25.80 | 0.0111 |
| | $^{208}$Tl | 583.2 | 85.00 | 0.0161 |
| $^{40}$K | - | 1460.8 | 10.66 | 0.0075 |

## 3.3. Lower Limit Detection

Lower Limit Detection (LLD) is a parameter related to the absence of a signal resulting from radiation. In high-resolution gamma-ray spectrometry, on the search for energy peaks, the sensitivity can be set using a threshold parameter which usually represents the level of significance in terms of standard deviation. (Blaauw, 2016; De Geer, 2004). In this study, it is necessary to measure LLD because environmental samples have relatively small activity, and there is a possibility that radioactive material will be enumerated other than samples such as background radiation which can affect the enumeration. LLD measurements are carried out by counting the background radiation, which will be used as a benchmark for radionuclide activity that is worthy of further review. The results of LLD measurements can be seen in Table 5.

**Table 5.** The results of the lower limit detection to determine the minimum limit of radionuclide activity so that it can be analyzed further

| Parent nuclide | Daughter nuclide | LLD (Bq) |
|---|---|---|
| $^{226}$Ra | $^{214}$Pb | 0.287 |
|  | $^{214}$Pb | 0.186 |
|  | $^{214}$Bi | 0.309 |
| $^{232}$Th | $^{212}$Pb | 0.097 |
|  | $^{228}$Ac | 0.344 |
|  | $^{228}$Ac | 0.292 |
|  | $^{208}$Tl | 0.081 |
| $^{40}$K | - | 3.849 |

## 3.4. Radioactivity of Soil and Coal Samples

Each cluster's soil and coal samples were counted using a gamma spectroscopy system. The counted results are in the form of an energy spectrum, and identification of the radionuclides contained in the sample can be carried out. Data on the concentration of radionuclides in each soil and coal sample are shown in Table 6 and Table 8, then the activity of each type of sample is compared with secondary data found in the literature or research elsewhere.

Based on Table 6, coal samples can have different concentrations in each cluster. This is because the mine pit and stockpile clusters have different coal types where the quantity of natural radionuclides varies greatly depending on the ash content and calorific value (Uslu & Gökmeşe, 2010). In addition, sampling does not pay attention to the type of coal taken but only to human activities, types of activities, geographical conditions and company permits.

**Table 6.** The results of the concentration of each radionuclide in the coal sample of the Tanjung Enim coal mine, Indonesia

| Sample Code* | Concentration (Bq/kg) | | |
|---|---|---|---|
|  | $^{226}$Ra | $^{232}$Th | $^{40}$K |
| SP-Bb-1 | 86.478 | 111.149 | 2174.479 |
| SP-Bb-2 | 61.096 | 86.888 | 1809.154 |
| SP-Bb-3 | 63.841 | 60.151 | 1602.686 |
| PT-Bb-1 | 79.044 | 102.495 | 1686.653 |
| PT-Bb-2 | 68.960 | 90.622 | 1651.146 |
| PT-Bb-3 | 75.388 | 70.127 | 1888.179 |
| Average | 72.468 | 86.905 | 1802.049 |

*Sp: *Stockpile*, PT: Mining pit, Bb: Coal

Most of the $^{232}$Th in coal is contained in phosphate minerals such as monazite or apatite. On the other hand, uranium and $^{226}$Ra are found in coal's mineral and organic fractions (Makolli et al., 2020). The high concentration of $^{40}$K in coal is due to bitumen rocks of plant origin having compounds that bind to potassium. Plants that are converted to coal use potassium as an essential element involved in functions such as nutrition, enzyme activation, osmotic regulation, growth, and plant development so that potassium levels in plants have a significant quantity, about 25% of the total mineral (Gupta et al., 1998; John et al., 2022).

The results of the calculation of the average concentration on coal samples from the Tanjung Enim's coal mine are for $^{226}$Ra of 72.468 Bq/kg, $^{232}$Th of 86.905 Bq/kg, and $^{40}$K of 1802.049 Bq/kg. Then the radionuclide concentration of the coal sample in this study was compared with the world average value reported by UNSCEAR and various studies that have been carried out in other coal mines, as shown in Table 7.

**Table 7.** Comparison of concentrations of $^{226}$Ra, $^{232}$Th, and $^{40}$K coal samples at Tanjung Enims locations from various coal samples at other research locations

| Location | Concentrations (Bq/kg) | | | Reference |
|---|---|---|---|---|
|  | $^{226}$Ra | $^{232}$Th | $^{40}$K |  |
| Tanjung Enim - Indonesia | 72.468 | 86.905 | 1802.049 | Present study |
| Kiwira - Tanzania | 448 | 455 | 3069 | (Makudi et al., 2018) |
| Bangladesh | 54.3 | 92.39 | 241.0 | (Habib et al., 2019) |
| Turki | 70 | 20 | 229 | (Akkurt et al., 2009) |
| Coorg - India | 10.46 | 66.37 | 426.77 | (Prakash et al., 2017) |
| Brazil | - | 122 | 1126 | (Hajj et al., 2017) |
| Swiss | - | 70 | 1005 | (Hajj et al., 2017) |
| Parana State - Brazil | 321 | 22 | 191 | (Flues et al., 2002) |
| Gombe - Nigeria | 8.18 | 6.97 | 27.38 | (Kolo et al., 2016) |
| India | 16.8 | 19.5 | 37.2 | (Sahu et al., 2014) |
| Albaha - Saudi Arabia | 35 | 31.52 | 843.63 | (Al-Zahrani, 2017) |
| UNSCEAR | 35 | 30 | 400 | (UNSCEAR, 2000) |

Based on Table 7, the concentration of $^{226}$Ra is 2 times greater, $^{232}$Th is 2.8 times greater, and $^{40}$K is 4.5 times more involved than the world average data provided by UNSCEAR 2000. On the other hand, the comparison of radionuclide concentrations Coal samples obtained by other studies shows that the concentration of $^{226}$Ra in coal samples tends to be higher than several studies that have been conducted in Bangladesh, India, Nigeria, and Saudi Arabia. In addition, the average activity of $^{226}$Ra is similar to the values reported in coal from Turkey and lower than the values reported from Tanzania and Brazil. Meanwhile, the concentrations of $^{232}$Th and $^{40}$K in the coal samples analysed in this study were relatively higher when compared to reports in several regions.

The radioactivity concentrations of the three radionuclides ($^{226}$Ra, $^{232}$Th, $^{40}$K) in the soil samples are presented in Table 8. Soil samples were collected from 2 different clusters in the coal mining environment, namely the landfill site or soil stockpile and viewpoint mining pit location. The two clusters are areas around coal mine pits where human activity is highest. This is because the landfill is an area that directly receives the results of non-coal mineral extraction during the mining process and at the viewpoint are posts adjacent to the mining process.

Based on Table 8, the same cluster of soil samples showed different concentrations. This is because natural radionuclides in soil and rock depend on soil type, mineral content, and geological conditions hile sampling, especially clusters of soil piles with various kinds of soil from mining excavations. This can significantly affect the distribution of radionuclides in the soil. The type of rock also determines the concentration of radionuclides present in it. Higher levels of radionuclides are often found in igneous rocks, such as granite, and lower levels are usually found in sedimentary rocks (Tzortzis et al., 2003).

**Table 8.** The results of the concentration of each radionuclide in the soil sample of the Tanjung Enim coal mine, Indonesia

| Sample Code* | Concentration (Bq/kg) | | |
|---|---|---|---|
| | $^{226}Ra$ | $^{232}Th$ | $^{40}K$ |
| TT-Th-1 | 73.326 | 107.138 | 1401.717 |
| TT-Th-2 | 71.816 | 85.788 | 1414.082 |
| TT-Th-3 | 82.674 | 98.791 | 1483.362 |
| VP-Th-1 | 98.085 | 117.039 | 1693.255 |
| VP-Th-2 | 77.967 | 89.284 | 1429.104 |
| VP-Th-3 | 71.361 | 103.213 | 1238.131 |
| Average | 79.205 | 100.209 | 1443.275 |

*TT: Landfill, VP: Viewpoint, Th: Soil

The results of the calculation of the average concentration in soil samples from the Tanjung Enim's coal mine are for $^{226}Ra$ of 79.205 Bq/Kg, $^{232}Th$ of 100.209 Bq/kg, and $^{40}K$ of 1443.275 Bq/kg. Then the soil sample concentrations were compared with the world average values reported by UNSCEAR and various studies carried out in other coal mines, shown in Table 9.

**Table 9.** Comparison of Concentrations of $^{226}Ra$, $^{232}Th$, and $^{40}K$ in soil samples at Tanjung Enims locations from various soil samples at other research locations

| Location | Concentration s(Bq/kg) | | | Reference |
|---|---|---|---|---|
| | $^{226}Ra$ | $^{232}Th$ | $^{40}K$ | |
| Tanjung Enim - Indonesia | 79.205 | 100.209 | 1443.275 | Present study |
| Kiwira - Tanzania | 378 | 331 | 2632 | (Makudi et al., 2018) |
| Villanueva - Kolombia | 44.25 | 62.8 | 1596.3 | (Salazar et al., 2021) |
| Tamil Nadu - India | - | 279.53 | 108.35 | (Ajithra et al., 2017) |
| Rajasthan - India | 50.28 | 34.16 | 587.45 | (Mehra et al., 2021) |
| Assuit - Mesir | 2670 | 1401 | 1495 | (El-Gamal et al., 2013) |
| Kutha - Iraq | 19.1565 | 54,501 | 179,578 | (Oleiwi, 2021) |
| Baoji - Cina | 40.3 | 59.6 | 749.7 | (Lu et al., 2012) |
| Orlu - Nigeria | - | 1.64 | 134.13 | (Mbonu & Ben, 2021) |
| Sanliurfa - Turki | 20.8 | 24.95 | 298.61 | (Bozkurt et al., 2007) |
| Guangyao - Cina | 26.8 | 8.87 | 453.81 | (Wang & Ye, 2021) |
| UNSCEAR | 32 | 45 | 420 | (UNSCEAR, 2000) |

Based on Table 9, the concentration of $^{226}Ra$ is 2.5 times greater, $^{232}Th$ is 2.2 times, and $^{40}K$ is 3.4 times greater than the concentrations recommended by UNSCEAR 2000 (UNSCEAR, 2000). Compared with previous studies, the elements $^{226}Ra$, $^{232}Th$, and $^{40}K$ were relatively higher than the results reported elsewhere but still below Tanzania and Egypt. The variation in the concentration of radionuclides comes from mining activities, but it is also caused by the regional geological conditions of the area where each region will have different properties.

The accumulation of natural radionuclides in the mining environment is relatively high for soil and coal samples. This is related to contamination from coal by-products that are also lifted to the ground surface during the coal mining process, which continues for an extended period. The build-up of by-products at a site can have an increased concentration of radioactive material and thus potentially produce radiological problems for mine workers.

*3.5. Hazard Index*

The hazard index of soil and coal samples can be calculated by calculating the radium equivalent activity parameters, internal hazard index, and external hazard index. The radium equivalent index can be interpreted that with various radionuclide concentrations for $^{226}Ra$, $^{232}Th$, and $^{40}K$ in each sample, it will be equivalent or equivalent to a concentration of elemental radium (Mohapatra et al., 2013; Wang & Ye, 2021). Other parameters such as the external hazard index describe the radiological hazard potential through the external exposure pathway received by mining workers due to the combination of natural radiation generated by $^{226}Ra$, $^{232}Th$, and $^{40}K$. The internal hazard index describes the potential radiological hazard due to exposure to internal radiation due to the entry of natural radionuclides ($^{26}Ra$, $^{232}Th$, and $^{40}K$) into the body of mining workers (Mehra et al., 2021). The calculations of hazard index are presented in Table 10.

**Table 10.** Hazard index parameters for each sample in the Tanjung Enim coal mine, Indonesia

| Sample Code | $Ra_{eq}$ (Bq/kg) | $H_{ex}$ | $H_{in}$ |
|---|---|---|---|
| SP-Bb-1 | 412.855 | 1.115 | 1.349 |
| SP-Bb-2 | 324.651 | 0.877 | 1.042 |
| SP-Bb-3 | 273.263 | 0.738 | 0.911 |
| PT-Bb-1 | 355.484 | 0.960 | 1.174 |
| PT-Bb-2 | 325.687 | 0.880 | 1.066 |
| PT-Bb-3 | 321.060 | 0.867 | 1.071 |
| TT-Th-1 | 334.465 | 0.903 | 1.101 |
| TT-Th-2 | 303.378 | 0.819 | 1.013 |
| TT-Th-3 | 338.164 | 0.913 | 1.137 |
| VP-Th-1 | 395.831 | 1.069 | 1.334 |
| VP-Th-2 | 315.684 | 0.853 | 1.063 |
| VP-Th-3 | 314.292 | 0.849 | 1.042 |
| Coal Sample Average | 335.500 | 0.906 | 1.102 |
| Soil Sample Average | 333.636 | 0.901 | 1.115 |
| UNSCEAR | ≤ 370 | ≤ 1 | ≤ 1 |

Based on Table 10, the coal sample has an average $Ra_{eq}$ value of 335.500 Bq/kg, $H_{ex}$ with an average of 0.906, and $H_{in}$ with an average of 1.102. While the soil sample has an average $Ra_{eq}$ value of 333.636 Bq/kg, $H_{ex}$ with an average of 0.901, and $H_{in}$ with an average of 1.115. Based on the hazard index calculation, the value of each index is compared with the recommendations given by UNSCEAR. The comparison shows that most $Ra_{eq}$ and $H_{ex}$ values are below the recommended value, while the $H_{in}$ shows the opposite, which exceeds the recommended limit ($H_{in}$ ≤ 1). Thus, the potential for radiological hazards affecting mining workers will be more significant through internal pathways. This can be due to dusty mining conditions and the possibility of radioactive particulates being lifted into the atmosphere, causing natural radionuclides to be inhaled when breathing or eaten and drinking while in the contaminant zone.

*3.6. Radiation Dose and Cancer Risk*

RESRAD-Onsite 7.2 simulation was used to calculate the dose and risk parameters of cancer received by coal mining workers due to an increase in the concentration of natural radionuclides. RESRAD-Onsite 7.2 is used because the cluster location is considered an area contaminated with radioactive material. The simulation was carried out by opening 5 paths, namely external gamma, inhalation, radon, soil ingestion, and drinking water. External gamma is defined as exposure resulting from contaminated soil to receptors standing on it. The inhalation route is a route of exposure to radionuclides that are inhaled into human respiratory organs. The radon pathway consists of 2, namely water independent radon (exposure produced by radon suspended in the air and water-dependent radon) and water-dependent water (exposure to radon dissolved in groundwater). Soil ingestion and drinking water pathways are radionuclide exposure pathways that result from being swallowed by soil or water into human digestive organs. (Argonne National Laboratory, 2018; Yu et al., 2001).

The input parameters used in the simulation are the specific activity average values of $^{226}$Ra, $^{232}$Th, and $^{40}$K for each cluster. The mining worker is assumed to be an adult male with a working period of 40 years, weight 70 kg, height 170 cm. Every year, workers will work in the contaminant zone for 8 hours per day (4 hours indoors and 4 hours outdoors), 6 days per week, 50 weeks per year.

The dose conversion factor parameter used in the simulation is based on Federal Guidance Report 12. Then the environmental transport factor will have a different value for each exposure path. These differences can be due to a path having specific environmental factors. Assuming the protective occupancy factor consists of 2 components, the indoor and outdoor fractions are 0.17 and 0.17, respectively. The contaminant form factor is assumed to be circular with no cover material. The area factor in the simulation uses the point-kernel method where this method will divide the contamination area into a small grid that contributes to the calculation of the radiation dose in the contaminant zone (Prokhorets et al., 2007). The assumptions of environmental parameters required for the simulation of RESRAD-Onsite 7.2 are presented in Table 11.

Table 11. The assumption of environmental parameters used to simulate the annual effective dose rate and excess lifetime cancer risk received by coal mine workers at Tanjung Enim, Indonesia with RESRAD-ONSITE 7.2 software

| Parameter | Value |
| --- | --- |
| Regulatory standards (Eckerman & Ryman, 1993) | Federal Guide Report 12 |
| Contamination thickness | 1 m |
| Cover thickness | 0 meter |
| Erosion rate (Noferiandani & Kironoto, 2008) | 0.00106 m/year |
| Wind speed | 1.562 m/s |
| Evaporation coefficient | 0.5 |
| Precipitation | 2.678 m/year |
| Irrigation (Yu et al., 1993) | 0.1 m/year |
| Irrigation type | Over head |
| Runoff coefficient | 0.4 |
| Soil ingestion (Oregon State University, 2011) | 73 g/year |
| Exposure duration | 40 years with 300 days/year or 2400 hours/year |
| Contamination zone shape | Circular |

Before the dose calculation simulation is carried out in each cluster, a variation of the contaminant thickness is carried out in one of the clusters to see which path has the most significant impact on the workers. The simulation results on variations in contaminant thickness in the stockpiled cluster are as shown in Table 12 and Figure 6.

Table 12. Simulation results of annual effective dose rates on 5 exposure pathways (external gamma, inhalation of dust, radon exposure, drinking water, and ingestion of soil) at various contaminant thickness at the Tanjung Enim coal mine, Indonesia

| Contaminated zone thickness (m) | Annual Effective Dose Rate (mSv/year) | | | | | |
| --- | --- | --- | --- | --- | --- | --- |
| | External Gamma | Inhalation | Radon (Water Independent) | Radon (Water Dependent) | Drinking Water | Soil Ingestion |
| 0.25 | 0.510915 | 0.008077 | 0.692257 | 1.044E-06 | 0.019052 | 0.000651 |
| 0.5 | 0.589744 | 0.008077 | 0.920803 | 1.044E-06 | 0.019183 | 0.000651 |
| 1 | 0.604790 | 0.008077 | 1.044134 | 1.044E-06 | 0.019248 | 0.000651 |
| 2 | 0.607407 | 0.008077 | 1.116948 | 1.044E-06 | 0.019337 | 0.000651 |

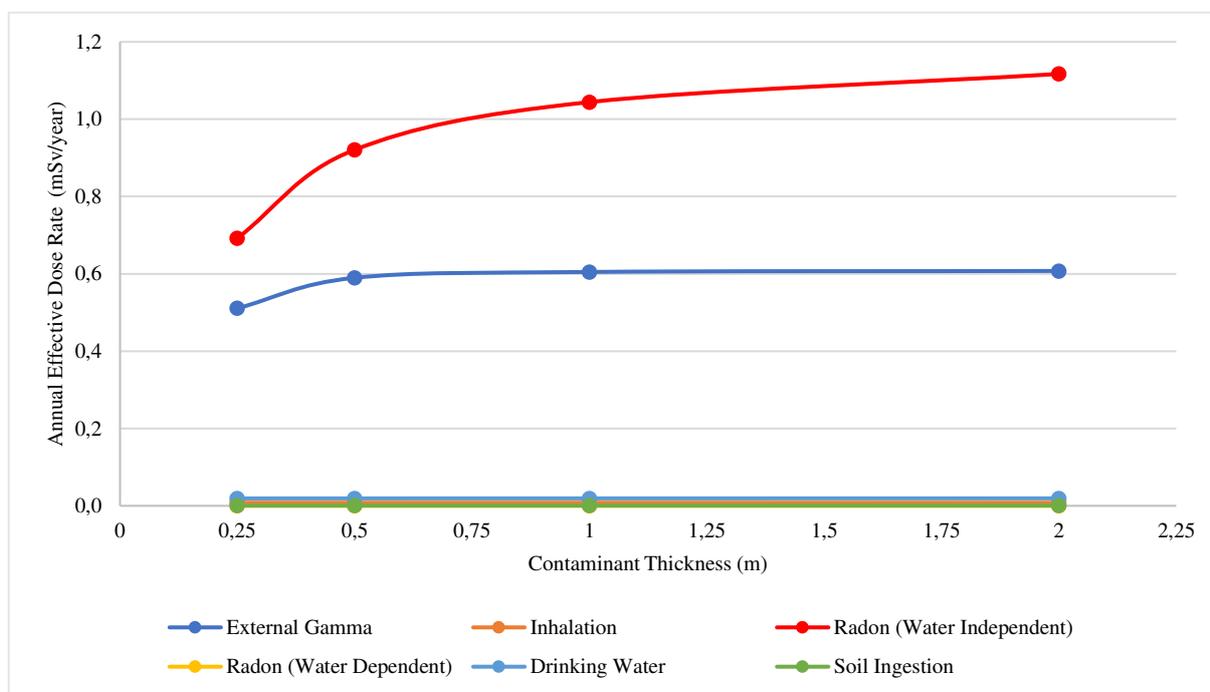

Figure 6. Prediction of average annual effective dose rate in stockpile clusters with variations in contaminant thickness

In the stockpile cluster, variations in the thickness of the contaminants were carried out to see the exposure path that had the most dominant effect on the dose rate. The most significant contribution to radiation exposure came from the radon water independent and external gamma pathways, while the water-dependent, inhalation, soil ingestion, and drinking water radon pathways were not very large. The immense contribution of water independent radon exposure due to hot and dusty mining conditions can make radon easily suspended in the air to provide radiation exposure to workers.

In general, by increasing the thickness of the contaminant, the resulting dose rate will be greater because of the build-up factor that causes radiation scattering in the soil to increase the exposure to mining workers. However, increasing contaminant thickness does not increase the dose rate produced by the external gamma path. The radiation emitted from the deepest soil will be attenuated by the material or the soil above it.

The radon exposure pathway (water-dependent) does not provide a significant dose rate because the accumulation of radium in the contaminant zone has dissolved with groundwater, making it difficult for these radionuclides to be suspended in the air. The inhalation route produces an insignificant dose rate due to radiation exposure that does not come from radon, the decay of radium or thorium, but from tritium, $^{14}C$ in $CO_2$ gas, and other radionuclides whose concentrations are deficient in the environment. Soil ingestion and drinking water pathways do not contribute a large dose because ingested radionuclides can have a biological half-life that allows radionuclides to be excreted from the body through the excretory system.

**Table 13.** Simulation results of annual effective dose rate and excess lifetime cancer risk received by coal mine workers at Tanjung Enim, Indonesia at a contaminant thickness of 1 meter

| Cluster | Pathway | Annual Effective Dose (mSv/year) | | ELCR ($10^{-3}$) |
|---|---|---|---|---|
| | | Value of each pathway | Total | |
| Coal Stockpile | External Gamma | 0.605 | 1.677 | 3.354 |
| | Inhalation | 0.008 | | |
| | Radon (Water Independent) | 1.044 | | |
| | Radon (Water Dependent) | 0.000 | | |
| | Soil Ingestion | 0.001 | | |
| | Drinking water | 0.019 | | |
| TSBC Mine Pit | External Gamma | 0.597 | 1.654 | 3.309 |
| | Inhalation | 0.008 | | |
| | Radon (Water Independent) | 1.030 | | |
| | Radon (Water Dependent) | 0.000 | | |
| | Soil Ingestion | 0.001 | | |
| | Drinking water | 0.019 | | |
| Soil stockpile | External Gamma | 0.572 | 1.587 | 3.174 |
| | Inhalation | 0.008 | | |
| | Radon (Water Independent) | 0.988 | | |
| | Radon (Water Dependent) | 0.000 | | |
| | Soil Ingestion | 0.001 | | |
| | Drinking water | 0.018 | | |
| Viewpoint TSBC | External Gamma | 0.552 | 1.532 | 3.063 |
| | Inhalation | 0.007 | | |
| | Radon (Water Independent) | 0.954 | | |
| | Radon (Water Dependent) | 0.000 | | |
| | Soil Ingestion | 0.001 | | |
| | Drinking water | 0.018 | | |
| Viewpoint Stockpile | External Gamma | 0.695 | 1.928 | 3.856 |
| | Inhalation | 0.009 | | |
| | Radon (Water Independent) | 1.201 | | |
| | Radon (Water Dependent) | 0.000 | | |
| | Soil Ingestion | 0.001 | | |
| | Drinking water | 0.022 | | |
| | Average | | 1.676 | 3.351 |

The simulation results for calculating the dose and cancer risk for each cluster using RESRAD-Onsite 7.2 are presented in Table 13. The total annual effective dose rate is the sum of all the doses generated from each pathway scenario. The annual effective dose rate and the mean ELCR were 1.676 mSv/year and 3.351×$10^{-3}$, respectively. The annual effective dose rate received by mining workers exceeds the dose limit for non-radiation workers as stipulated in ICRP publication 127 and publication 103, which is below 1 mSv/year due to the accumulation of TENORM (Ojar et al., 2014; US EPA, 2000). In addition, the risk of cancer or ELCR shows a high value, even up to 14 times greater than that recommended by UNSCEAR, which is 2.4×$10^{-4}$. This ELCR represents the risk of developing cancer for individuals or workers who will spend most of their life in the studied area so that intervention measures are needed to prevent long-term radiological hazards.

Based on the results, an intervention is needed to limit the dose rate and long-term radiological risks coal mining workers receive. The author recommends that mineworkers not be in the contaminant zone more than 4,8 hours/day or 4 days/week for not to exceed the dose limit, which is 1 mSv/year. Besides that, interventions can be carried out by using additional personal protective equipment such as P100 masks to minimize radon entry into the internal pathway (Gardner et al., 2013). The intervention by using a mask is expected not to shorten the duration while in the contaminant zone and maintain productivity in coal mines. Other intervention actions that can be taken are to inform employees about the nature and level of risk from the accumulation of radon and other TENORM materials and minimise the use of water from contaminant locations that are used for eating and drinking.

## 4. Conclusions

The identification results of radionuclides contained in the soil and coal samples were $^{226}Ra$ with its decay products ($^{214}Pb$ and $^{214}Bi$), $^{232}Th$ with its decay products ($^{212}Pb$, $^{228}Ac$, and $^{208}Tl$), and $^{40}K$. The coal samples have the mean concentrations for $^{226}Ra$, $^{232}Th$, and $^{40}K$, respectively, 72.468 Bq/kg, 86.905 Bq/kg, and 1802.049 Bq/kg. Meanwhile, the mean concentrations for $^{226}Ra$, $^{232}Th$, and $^{40}K$ in the soil samples were 79.205 Bq/kg, 100.209 Bq/kg, and 1443.275 Bq/kg, respectively.

Hazard indexes such as $Ra_{eq}$, $H_{in}$, and $H_{ex}$ in coal samples have the mean values of 335.500 Bq/kg, 1.102, 0.906, respectively. While the soil samples have the average $Ra_{eq}$, $H_{in}$, and $H_{ex}$ values of 333.636 Bq/kg, 1.115, and 0.901, respectively. The $Ra_{eq}$ and $H_{ex}$ parameters have met the UNSCEAR recommendations, namely $Ra_{eq}$ 370 Bq/kg and $H_{ex} \leq 1$, while the $H_{in}$ value is more than 1 which creates a potential internal radiological hazard.

The total annual effective dose rate received by miners based on the 5 RESRAD-Onsite 7.2 pathways namely external gamma, inhalation, radon, soil ingestion, and drinking water, is 1.676 mSv/year. Meanwhile, the ELCR received by mining workers due to exposure to accumulated natural radiation is $3.351 \times 10^{-3}$. The ELCR value obtained is 14 times greater than the recommendation given by UNSCEAR, which is $2.4 \times 10^{-4}$.

Intervention at the Tanjung Enim's coal mine is needed to reduce the radiological hazards. The intervention recommendations are workers not exceeding 4.8 hours/day or 4 days/week in the contaminated zone. An intervention that does not reduce working hours is to wear a mask, such as P100, while in the contaminant zone. Another intervention is to inform employees about the nature and level of risk from the accumulation of radon and other TENORM materials and minimise the use of water from contaminant locations for consumption.

## Acknowledgements

The authors would like to thank the Department of Nuclear Engineering and Physics Engineering Gadjah Mada University, Yogyakarta National Research and Innovation Agency, PT. Bukit Asam Tbk. and all of those who contributed to acquiring research material.

## Data Availability

The datasets generated during and/or analysed during the current study are available from the corresponding author on reasonable request.

## Declaration of Competing Interest

The authors declare that they have no known competing financial interests or personal relationships that could have appeared to influence the work reported in this paper.